\documentclass[usenatbib]{mn2e}
\usepackage{color} 

\usepackage{graphics}
\usepackage{epsfig}
\usepackage{natbib}

\begin{document}

\title[Galaxies with Red Mid-Infrared Colours]
{The Physical Properties of Galaxies with Unusually Red Mid-Infrared Colours}

\author [G.Kauffmann] {Guinevere Kauffmann\thanks{E-mail: gamk@mpa-garching.mpg.de}\\
Max-Planck Institut f\"{u}r Astrophysik, 85741 Garching, Germany}

\maketitle

\begin{abstract} 
The goal of this paper is to investigate the physical nature of galaxies
in the redshift range $0.02<z<0.15$
that have strong excess emission at mid-IR wavelengths and to determine
whether they host a population of accreting black holes that cannot be
identified using optical emission lines.  We show that
at fixed stellar mass $M_*$ and $D_n(4000)$, the distribution of 
[3.4]-[4.6] $\mu$m (WISE W1-W2 band)
colours is sharply peaked, with a  
long tail to much redder W1-W2 colours.  We
introduce a procedure to pull out the red outlier population based
on a combination of three stellar population diagnostics. When compared
with optically-selected AGN, red outliers are more likely to  
be found in massive galaxies, and they tend to have lower stellar mass densities,   
younger stellar ages and higher dust content than optically-selected AGN hosts. 
They are twice as likely
to be detected at radio wavelengths. 
We examine W1-W2 colour profiles for
a subset of the nearest, reddest outliers and find that most are not
centrally peaked, indicating that the hot dust emission is spread
throughout the galaxy.  We find that radio luminosity is the quantity that
is most predictive of a redder central W1-W2 colour. Radio-loud galaxies
with centrally concentrated hot dust emission are almost always
morphologically disturbed, with compact, unresolved emission at 1.4 Ghz.
Eighty percent of such systems are identifiable as  AGN using
optical emission line diagnostics.
\end{abstract}


\section {Introduction} 
Our understanding of the co-evolution of
supermassive black holes and their surrounding host galaxies has benefited
greatly over the last 15 years as a result of many large imaging and
spectroscopic surveys that have characterized both the emission from matter
falling into black holes and from stars and ionized gas in the host galaxy
over a wide range in cosmic time (e.g. Kauffmann et al 2003; Ueda et al
2003; Croom et al 2004; Richards et al 2006). A broad theoretical picture
has now emerged that the largest black holes in the Universe with masses in
excess of $10^9 M_{\odot}$ are already assembling in quasars at redshifts
$\sim 6$, and that by the present-day such black holes are found in giant
elliptical galaxies located in massive groups or clusters (e.g.  Volonteri
et al 2003; Granato et al 2004; Hopkins et al 2008).  The present-day
accretion rates onto black holes in such enviroments are believed to be
low, with most of the emitted energy redirected outwards in the form of
relativistic jets that emit at radio wavelengths (e.g. Croton et al 2006;
Somerville et al 2008; Sijacki et al 2015). These are the FRI radio
galaxies that can also be identified optically through their low-ionization
emission lines.  At the present day, black holes with masses of $\sim 10^6
M_{\odot}$ comparable to the central object in our own Milky Way are still assembling in
the bulges of spiral galaxies and these make up the majority of the Seyfert
population (Marconi et al 2004; Heckman et al 2004).

The conditions under which black holes are able to form and grow in mass is
still a subject of controversy. Prompted by theoretical arguments and
simulated scenarios of black hole growth, much attention has focused on the
role of galaxy mergers and interactions as a mechanism that can cause  gas
to lose angular momentum, flow inwards to the very central regions of the
galaxy and accrete onto the central supermassive black hole (Heckman et al
1986; Sanders et al 1988;  Kauffmann \& Haehnelt  2000; Di Matteo, Springel
\& Hernquist 2005).  This mechanism is observed 
to induce powerful
bursts of star formation in galaxy centers (see Sanders \& Mirabel 1996 for
a review), but the degree to which the black hole will grow under such
circumstances is still poorly understood.
The amount of growth will likely depend on how much gas is driven out by energetic
processes such as supernova or accretion-driven winds.

The so-called Soltan argument (Soltan 1982) relates the energy released in
the optical region of the electromagnetic spectrum to the total mass
density in supermassive black holes at the present-day, showing that the
two are in reasonably good agreement. No clear connection between AGN
visible at optical wavelengths and galaxy mergers has been established
observationally. Studies that use irregularities in galaxy morphology or
the number of close neighbours as a probe show that once the amount of star
formation in the host is accounted for, there is no further boost in 
either of these quantities 
associated with emission from the central black hole (Li et al 2008;
Reichard et al 2009). The same result seems to hold for samples of AGN
selected from the largest X-ray imaging surveys (Cisternas et al 2011;
Kocevski et al 2012).

Nevertheless, the question still persists as to whether a merger-induced
phase of black hole growth may be missing from optical and X-ray samples
because it takes place in very dense and dusty environments, causing the
emission to be ``obscured''. Hard X-ray photons with energies above a
few keV travel through dust without
significant obscuration. There are now indications that some AGN
detected using data from the Nuclear Spectroscopic
Telescope Array (NuSTAR) (Harrison et al 2013) in the highest energy 3-24 keV band
are not easily identifiable using optical
emission line diagnostics (Chen et al 2017). However, such samples   
are still small and it is not yet possible to address the merger
connection.

Another more indirect way to identify an obscured phase of black hole
growth is to search for emission from  dust that has been  heated to extremely 
high temperatures by radiation from an accreting black hole. 
The black body spectrum from the atmospheres of stars peaks at 1.6
$\mu$m.  Emission from dust heated by young stars in strongly star-forming
galaxies peaks at 60 $\mu$m in the far-infrared. The mid-infrared is
therefore the electromagnetic window of choice for finding systems where
dust has been heated by radiation emitted by accreting black holes, which
has an spectrum peaked towards shorter wavelengths 
than the radiation emitted by young stars. Stern et
al (2005) introduced a wedge in the space of [3.6]-[4.5] $\mu$m versus
[5.8]-[8.0] $\mu$m IRAC colours where AGN could be identified. This
analysis made use of Spitzer Space Telescope data for  10,000  galaxies that
could be separated into AGN and normal galaxies based on their optical emission
line properties.  Subsequent work by Eckart et al (2010) showed that two
thirds of  X-ray selected AGN and almost all X-ray AGN with high ionization
emission lines  were also  located in this wedge in mid-IR colour space.

The launch of the Wide-field Infrared Survey Explorer (WISE) satellite in
2009 has  provided sensitive photometry of many millions of sources over the
entire sky in four mid-IR bands centred at 3.4, 4.6, 12 and 22 $\mu$m
(referred to as W1, W2, W3 and W4). AGN selection using WISE photometry was
addressed in  Stern et al (2012).  As illustrated in figures 1 and 2 of
that paper, ``pure'' AGN (i.e. AGN where no light from the host galaxy is
present) can be identified by means of a simple  W1-W2 colour cut (W1-W2$>$
0.8) at all redshifts. At redshifts greater than 1.5, the same cut is also
able to identify AGN in which the host galaxy contributes up to half the
emitted light. Most subsequent studies of high redshift AGN have therefore
adopted this simple cut in W1-W2 colour.

Low redshift studies have thus far used optical emission
line ratio diagnostics to identify AGN
(Rosario et al 2013; Shao et al 2013).  More recently, attempts have been
made to quantify the location of AGN in WISE mid-IR colour-colour or
colour-magnitude space using hard X-ray selected  samples
(Mateos et al 2012) using the principle that AGN should have red power-law
SEDs. Mateos et al show that this approach works well for the highest
luminosity AGN, but becomes incomplete at lower luminosities.
So far, the question of the physical nature of  missing modes of black hole accretion
has been addressed very sketchily in the literature.  

We note that the methodology discussed above {\em assumes no prior
information on the nature of the underlying host galaxy}. In this paper we
show that the W1-W2 colours of the normal galaxy population are clearly correlated with optical
diagnostics of their mean stellar age such as the 4000 Angstrom break index
D$_n$(4000), the H$\delta_A$ Balmer absorption line strength and the
specific star formation rate derived from the extinction-corrected
H$\alpha$ luminosity. When these correlations are defined using very large
galaxy and AGN  samples provided by the Sloan Digital Sky Survey, a clear
``main sequence'' of galaxies is visible in optical/mid-IR correlation
plots, as well as tail of outliers to very red $W1-W2$ colours. It is
the physical nature of this population of outliers that forms the main
subject matter of this paper.

In section 2, we describe how our sample of nearby galaxies is constructed
as well as our technique for selecting the red W1-W2 population. In
section 3, we analyze how galaxy properties such as  
stellar mass, stellar mass surface density, concentration index, specific
star formation rate, mean stellar age and radio luminosity
at 1.4 GHz compare to AGN selected by optical emission
line ratio diagnostics. In section 4, we analyze the structure of the hot dust in
these objects. We select a subset of galaxies with the reddest W1-W2 colours and
we investigate the extent to which the excess mid-IR emission is centrally
peaked, as would be expected if the hot dust is located in a central
torus-like structure. In Section 5 we summarize and discuss our results.

\section {Sample Construction} 
We begin with a magnitude-limited sample
constructed from the final data release (DR7; Abazajian et al. 2009) of the
Sloan Digital Sky Survey (York et al. 2000), which is described in Li et al.
(2012).  This sample contains 533,731 galaxies located in the main
contiguous area of the survey in the northern Galactic cap, with $r <
17.6$, $-24< M_r < -16$ and spectroscopically measured redshifts in the
range $0.001<z<0.5$. 87,060 of these are classified
as optical AGN based on their [OIII]/H$\beta$ versus 
[NII]/H$\alpha$ emission line ratios (Kauffmann et al. 2003). Here $r$ is the $r$-band Petrosian apparent magnitude,
corrected for Galactic extinction, and $M_r$ is the $r$-band Petrosian
absolute magnitude, corrected for evolution and K-corrected to its value at
$z=0.1$.  The apparent magnitude limit is chosen in order to select a
sample that is uniform and complete over the entire area of the survey. 
The median redshift of this sample is $z=0.088$,
with 10\% of the galaxies below z=0.033 and 10\% above z=0.16. In this
paper, we use galaxy properties available from the MPA/JHU database of
spectrum measurements (Brinchmann et al 2004) such as stellar masses and
star formation rates (SFRs), as well as optical AGN classifications based
on emission line flux measurements.

We cross-correlate this sample with the AllWISE Source Catalog, which
contains astrometry and photometry for 747,634,026 objects detected on the
deep AllWISE Atlas Intensity Images. The AllWISE program builds upon the
work of the successful Wide-field Infrared Survey Explorer mission (WISE;
Wright et al. 2010) by combining data from the WISE cryogenic and NEOWISE
(Mainzer et al. 2011) post-cryogenic survey phases to form the most
comprehensive view of the full mid-infrared sky currently available.  As
well as total magnitudes in 4 bands:  3.4, 4.6, 12, and 22 $\mu$m (W1, W2,
W3, W4), circular aperture photometry is available for 8 standard annulus
radii. For the W1,W2 and W3 bands, the radii are 5.5, 8.25, 11.0, 13.75,
16.5, 19.25, 22 and 24.75 arsec, respectively. We note that the SDSS
spectra are obtained through 3 arcsec diameter fibre apertures. When
correlating 4000 \AA\ break strengths with mid-IR colours, we use the 5.5 arcsec
circular aperture quantities. 533,612 galaxies are detected in both the W1
and W2 bands, i.e. 99.97\% of the original sample. The typical
error on the W1-W2 colours are a few hundredths of a magnitude, i.e. they
are very well measured.

In Figure 1, we plot galaxies in two stellar mass bins in the plane of
W1-W2 colour versus W2 magnitude. The greyscale contours indicate the
fraction of galaxies in the given bin. To calculate these fractions, we
weight each galaxy by $1/V_{max}$ where $V_{max}$ is the survey volume over
which the galaxy can be detected. The  contours are spaced logarithmically
and run from log F =-2.9 (black) to log F =-0.9 (white). We have subdivided
each sample into 4 different bins in 4000 \AA\ break strength $D_n(4000)$,
which are overplotted as different coloured line contours
(red, 1.8$<D_n(4000)<$2.0; yellow,
1.6$<D_n(4000)<$1.8; blue, 1.4$<D_n(4000)<$1.6; cyan, 1.2$<D_n(4000)<$1.4).
As can be seen, galaxies occupy a fairly tight locus in the W1-W2 versus W2
plane, with more luminous galaxies exhibiting redder colours.  The
different D$_n$(4000) subsamples are arranged in a well-ordered sequence,
with galaxies in  the lowest D$_n$(4000) bin  with the  youngest stellar
populations, exhibiting the reddest W1-W2 colours and brightest W2 magnitudes.

\begin{figure}
\includegraphics[width=90mm]{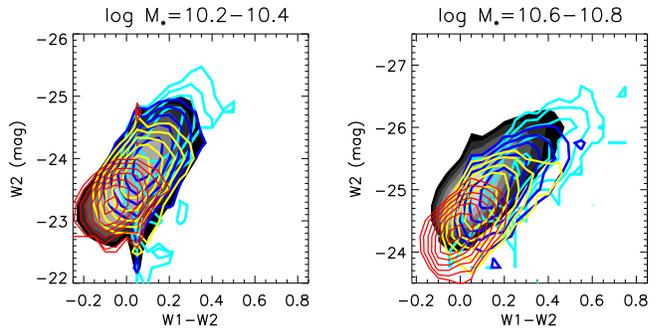}
\caption{ Galaxies in two narrow bins in stellar mass in the plane of W1-W2 colour
versus W2 magnitude. The greyscale contours indicate the logarithm of the
fraction of galaxies in the given bin and run from log F =-2.9 (black) to log F =-0.9 (white).  
We have subdivided
each sample into 4 different bins in 4000 \AA\ break strength $D_n(4000)$,
which are overplotted as different coloured line contours
(red, 1.8$<D_n(4000)<$2.0; yellow,
1.6$<D_n(4000)<$1.8; blue, 1.4$<D_n(4000)<$1.6; cyan, 1.2$<D_n(4000)<$1.4).
The coloured contours have the same spacing in log F as the black-and-white ones.
\label{models}}
\end{figure}

In Figure 2, we overplot the location of optically identified AGN in W1-W2
versus W2 magnitude plane. Results are shown in 4 different bins of stellar
mass.  AGN have been selected using the cut in the [NII]/H$\alpha$ versus
[OIII]/H$\beta$ plane given in Kauffmann et al (2003). As can be seen, in
the lowest mass bin, optically-selected AGN are biased towards bluer W1-W2
colours and fainter W2 luminosities compared to the sample as a whole. The
effect gets smaller at higher stellar masses. As shown in Figure 1,
galaxies with redder W1-W2 colours have systematically younger stellar
populations. This raises the question whether galaxies with actively
accreting black holes are systematically missing from the red part of the
W1-W2 versus W2 colour-magnitude plane. To answer this question, we need to
develop a method for identifying a subset of galaxies with red W1-W2
colours that are possible AGN candidates.

\begin{figure}
\includegraphics[width=90mm]{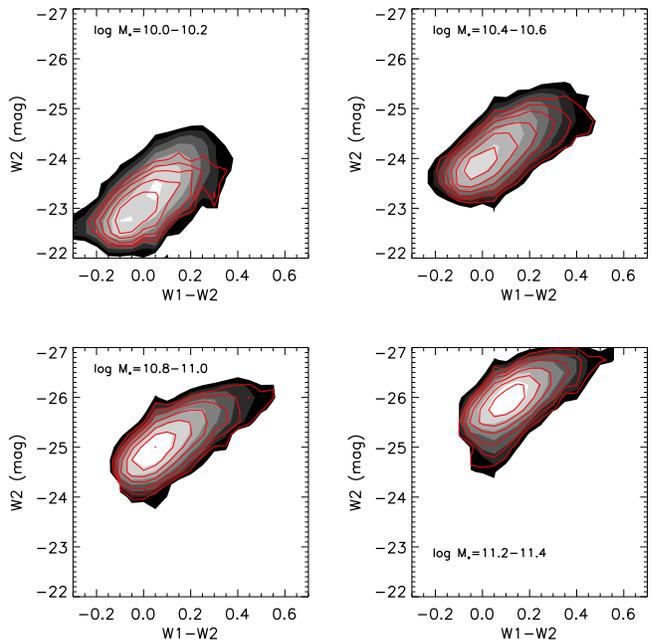}
\caption{Galaxies in four  narrow bins in stellar mass in the plane of W1-W2 colour
versus W2 magnitude. As in Figure 1, the greyscale contours indicate the logarithm of the
fraction of galaxies in the given bin.  
Results for the optical AGN sample in the same stellar mass range 
are overplotted as red contour lines.  
There are 413,969 galaxies in the stellar mass range
$10.0 < \log M_* < 11.4$, of which 80,974 are classified as optical AGN.
\label{models}}
\end{figure}

Figures 3 and 4 illustrate our methodology for AGN identification. In
Figure 3, we plot W1-W2 as a function of the logarithm of
the specific star formation rate  (log SFR/M$_*$) measured within the 
SDSS fibre aperture  for galaxies in a
narrow stellar mass range ($10.4<\log M_*<10.6$).  The star formation rates
are taken from the the MPA/JHU database (see Brinchmann et al (2004) for a detailed
description of the methodology). Note that we show only those galaxies
for which  both the H$\alpha$ and H$\beta$ emission lines
can be measured with reasonably high signal-to-noise 
($S/N>3$ in the H$\beta$ line), so that the SFR estimates
are based on emission line measurements rather than the 4000 \AA\ break itself.
The points have been colour coded
according to D$_n$(4000) value as indicated in the figure caption. The
majority of points occupy a narrow locus in the plots, but there is a
subset that scatter to much higher W1-W2 values at all values of log
SFR$/M_*$.  Figure 4 shows a similar plot of W1-W2 as a function of
H$\delta_A$ for galaxies in the same mass range, with galaxies colour-coded
the same way. Once again, the same conclusion holds as a function of
H$\delta_A$. Based on these two figures, we devise a method for identifying
red outliers as follows. The most accurately determined star formation rates based
on detections of all four of the strongest emission lines (H$\alpha$,
[OIII], H$\beta$ and [NII]) are generally available for galaxies with
specific star formation rates $\log SFR/M_* > -11$. For galaxies with lower
values of SFR/$M_*$, one or more of the emission lines are generally not
detected and the SFR estimates become increasingly inaccurate. We thus
identify outliers in  SFR/$M_*$ versus D$_n$(4000) space for galaxies with $\log
SFR/M_* > -11$ and in  H$\delta_A$ versus D$_n$(4000) space for galaxies with lower
specific star formation rates. We bin up the two planes in intervals of 0.15 in
D$_n$(4000), 0.25 in $\log$ SFR/$M_*$ and 0.1 in H$\delta_A$ and calculate
the distribution of W1-W2 colours in each bin. In subsequent plots, we
adopt two different definitions for ``outliers'' corresponding to the
upper 95th and 97th percentile points of the distribution.

\begin{figure}
\includegraphics[width=83mm]{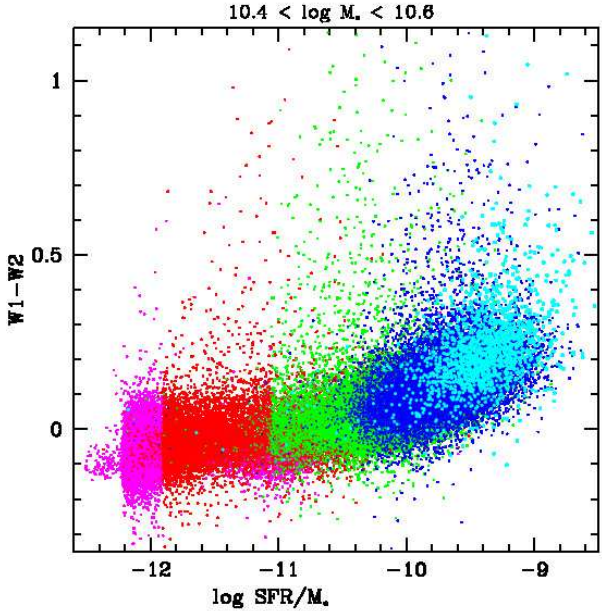}
\caption{ W1-W2 is plotted as a function of log SFR/M$_*$  for galaxies in a
narrow stellar mass range ($10.4<\log M_*<10.6$). The star formation rates
are based on extinction-corrected H$\alpha$ luminosities and are taken from
the MPA/JHU database. The points have been colour coded
according to D$_n$(4000) value as follows: 
(magenta, 2.0$<D_n(4000)<$2.2; red, 1.8$<D_n(4000)<$2.0; green,
1.6$<D_n(4000)<$1.8; blue, 1.4$<D_n(4000)<$1.6; cyan, 1.2$<D_n(4000)<$1.4).
\label{models}}
\end{figure}

\begin{figure}
\includegraphics[width=83mm]{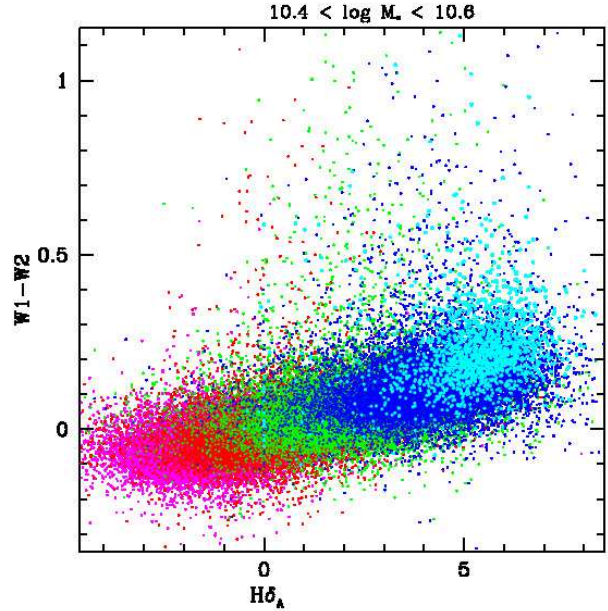}
\caption{ W1-W2 is plotted as a function of H$\delta_A$  for galaxies in a
narrow stellar mass range ($10.4<\log M_*<10.6$). 
The points have been colour-coded by D$_n$(4000) value as in Figure 3.
\label{models}}
\end{figure}

\section {Properties of the red W1-W2 outliers} 
We restrict our analysis
to galaxies in the mass range $\log M_* > 10.1$. Kauffmann et al (2003) show
that the fraction of optically selected AGN drops very strongly at stellar
masses lower than this. There have been suggestions that ``missed''
accreting black holes may be present in low mass galaxies from both hard
X-ray and mid-IR studies. We will not address this in this paper. Figure 5
compares the distribution of optically-selected AGN (black histograms) and
red W1-W2 outliers (red histograms) for a variety of galaxy properties. In
the case of the W1-W2 outliers, solid red histograms show results for the
95th percentile outliers, and dotted red histograms for the 97th percentile
cases.

\begin{figure}
\includegraphics[width=93mm]{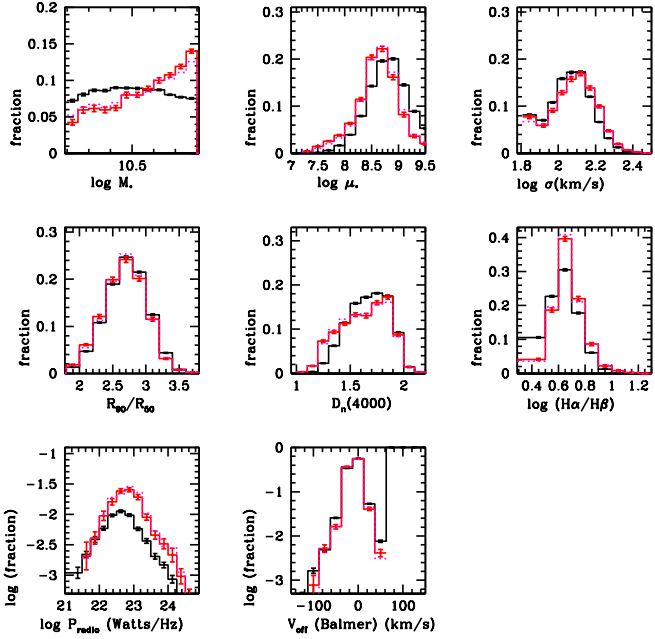}
\caption{ The distribution of optically-selected AGN (black histograms) and
red W1-W2 outliers (red histograms) are compared for a variety of galaxy properties:
(from left to right, and from top to bottom: log stellar mass, log stellar mass surface density
central stellar velocity dispersion, concentration index, 4000 \AA\ break strength,
logarithm of the Balmer decreement, log radio luminosity, and offset between the measured
velocity centroids of the Balmer emission lines and the systemic velocity of the
galaxy. In
the case of the W1-W2 outliers, solid red histograms show results for the
95th percentile outliers, and dotted red histograms for the 97th percentile
cases. 
\label{models}}
\end{figure}

As can be seen, the median stellar masses of the optical AGN and the red
outlier samples are similar, but the fraction of red outliers rises
as a function of stellar mass, whereas the fraction of optical AGN is
independent of stellar mass.  Red outliers are shifted towards slightly
lower stellar surface densities than optical AGN, but have
similar central stellar velocity dispersions and
concentration indices.  
As expected, the W1-W2 outliers have
younger stellar populations and higher dust extinctions than the optically
selected AGN.

In the bottom left panel we show the fraction of galaxies detected at
different radio luminosities, which were obtained by cross-correlating our
sample with VLA Faint Images of the Radio Sky at Twenty Centimeters source
catalogue (Becker, White \& Helfand 1995) using methodology
similar to that described in Best et al. (2005).
Interestingly, there is a factor 2 difference in radio-loud
fraction at radio luminosities greater than $10^{22.5}$ Watts Hz$^{-1}$.
We will come back to this
result later in the paper. In the bottom right panel, we compare the
distribution of shifts in the velocities of the Balmer emission lines with
respect to the systemic velocity of the galaxy. These velocity shifts are
from the MPA/JHU catalogue of spectrum measurements. We note that during
the line-fitting process, all the Balmer lines are fit simultaneously.
There is no significant difference between red outliers and optical AGN. 

In Figure 6, we compare two dimensional distributions of galaxy properties
for the two samples. The greyscale contours show the location of optically
selected AGN, while the red contours show the location of the red W1-W2
outliers in the 95th percentile sample.  The bottom left panel shows the
location of the two populations in the H$\delta_A$ versus D$_n$(4000)
plane. 
The top two panels show
the relation between D$_n$(4000) and two structural parameters, the stellar
surface mass density $\mu_*$ and the concentration parameter R90/R50 (R90
and R50 are the radii enclosing 90\% and 50\% of the $r$-band light from
the galaxy). The bottom right panel shows the distribution of Balmer
decrement as a function of stellar surface mass density. We see that the 
red outliers are slightly displaced to lower D$_n$(4000), higher H$\delta_A$ and
higher H$\alpha$/H$\beta$
values at fixed $\mu_*$, indicating that their stellar populations are
younger. We note, however, that the differences are bewtween the two samples
are not large.   

\begin{figure}
\includegraphics[width=87mm]{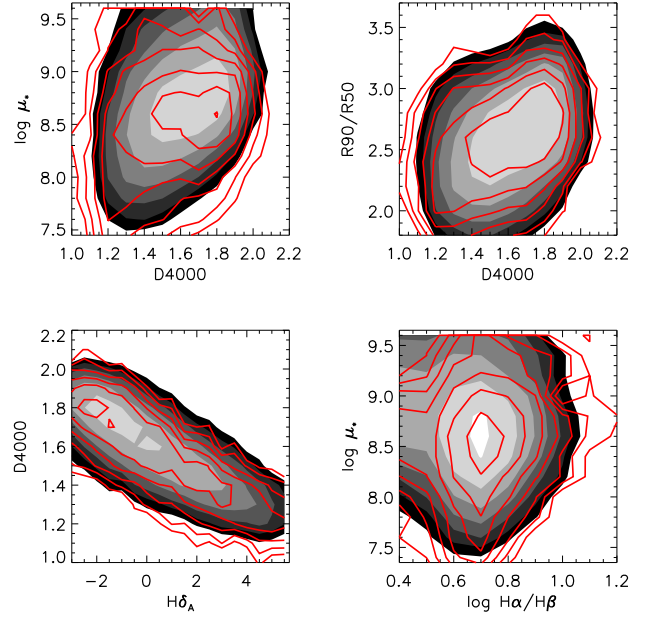}
\caption{ Two dimensional distributions of galaxy properties
for the two samples. The greyscale contours show the location of optically
selected AGN, while the red contours show the location of the red W1-W2
outliers in the 95th percentile sample.  
The top two panels show
the relation between D$_n$(4000) and two structural parameters, the stellar
surface mass density $\mu_*$ and the concentration parameter R90/R50 (R90
and R50 are the radii enclosing 90\% and 50\% of the $r$-band light from
the galaxy). 
The bottom left panel shows the
location of the two populations in the H$\delta_A$ versus D$_n$(4000)
plane. 
The bottom right panel shows the distribution of Balmer
decrement as a function of stellar surface mass density. 
\label{models}}
\end{figure}

The adoption of the 95th or 97th percentile points as the adopted cut on
the red outlier population is something of an arbitrary choice. In Figure
7, we investigate how galaxy properties change as W1-W2 colours become
redder.  We plot stellar mass, stellar mass density, concentration
parameter, velocity dispersion, D$_n$(4000) and Balmer decrement as a
function of the difference between the W1-W2 colour and the colour that
delineates the 95th percentile cut.  Each plotted point corresponds to a
bin containing a fixed number (200) galaxies, so the noise due to Poisson
sampling of the underlying distribution remains constant in each diagram.
Black points show the median of the distribution, red points the upper 75th
percentile of the distribution and blue points the lower 25th percentile.
Stellar mass and velocity dispersion decrease as a function of colour
difference , the concentration parameter remains roughly constant, and
stellar surface density and Balmer decrement increase.  The biggest change
as a function of $\Delta$(W1-W2) is the drop in D$_n$(4000).  We note that
almost all quantities no longer change significantly at $\Delta$(W1-W2)
values greater than about 0.15-0.2, suggesting that this delineates the
regime where stellar light from the host galaxy no longer contributes
significantly to the W1-W2 colour. In the next section, we investigate the
nature of this subset of the very reddest outliers in more detail.

\begin{figure}
\includegraphics[width=94mm, height=66mm]{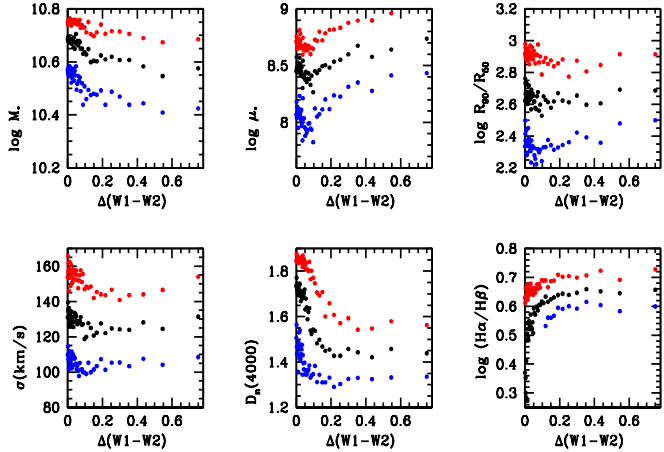}
\caption{Stellar mass, stellar mass density, concentration
parameter, velocity dispersion, D$_n$(4000) and Balmer decrement as a
function of the difference between the W1-W2 colour and the colour that
delineates the 95th percentile cut.  Each plotted point corresponds to a
bin containing a fixed number (200) galaxies, so the noise due to Poisson
sampling of the underlying distribution remains constant in each diagram.
Black points show the median of the distribution, red points the upper 75th
percentile of the distribution and blue points the lower 25th percentile. 
\label{models}}
\end{figure}

\section{Physical properties of the red outliers}

\subsection{Mid-IR colour profiles of the largest nearby red outliers}

We first investigate spatially resolved  W1-W2 colour profiles for a subset
of the nearest galaxies in our sample with the reddest W1-W2 colours. 
Based on the results shown in Figure
7, we select galaxies with $\Delta W1-W2 > 0.28$, redshifts $z < 0.035$,
and $r$-band Petrosian radii of 10 arcseconds or more.
 This results in a sample of 19 objects. SDSS cut-out images
are shown in Figure 8. As can be seen, most of the galaxies are
disk-dominated with red colours. Two out of the 19 appear to be merging
systems. One object is an extremely blue irregular galaxy and the light
within the central aperture is dominated by a bright star cluster.

\begin{figure}
\includegraphics[width=86mm]{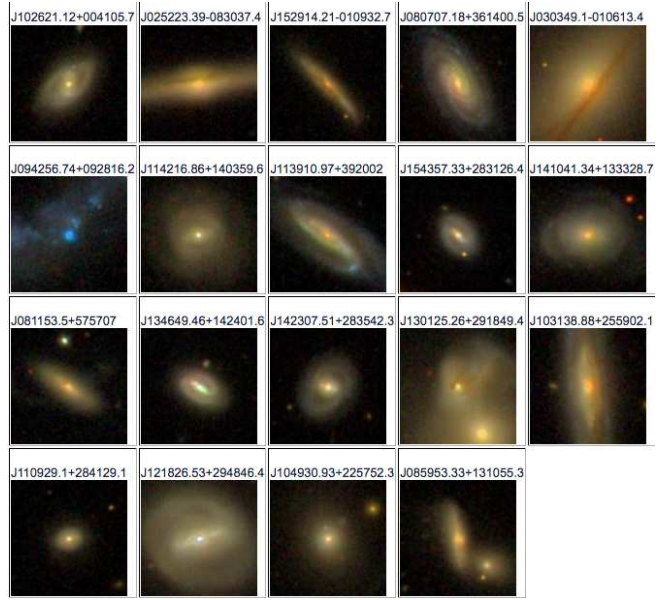}
\caption{ SDSS g,r,i colour cut-out images for a subset of the
nearest galaxies in our sample with the reddest W1-W2 colours (see text).
The orientation of the images is the same as in Figure 12 and so is the scale.
\label{models}}
\end{figure}

\begin{figure}
\includegraphics[width=99mm]{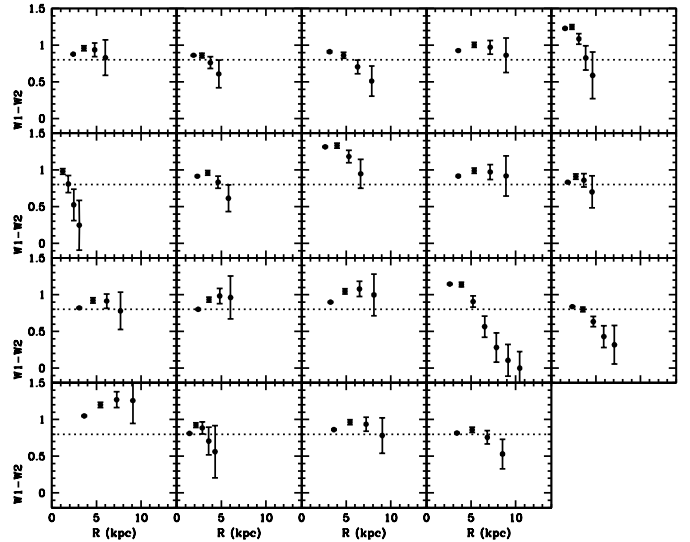}
\caption{ W1-W2 colour profiles for the same 19 galaxies shown in
Figure 8. The dotted line in each panel marks a W1-W2 colour of 0.8,
which was proposed by Stern et al. (2013) as a safe demarcation for
selecting AGN. The colours are calculated within rings corresponding
to the standard apertures used for WISE photometry, i.e. 0-5.5, 5.5-8.25, 
8.25-11.0, 11.0-13.75,
13.75-16.5, 16.5-19.25, 19.25-22 and 22-24.75 arsec.  
\label{models}}
\end{figure}

Figure 9 shows W1-W2 colour profiles for the same 19 galaxies shown in
Figure 8. The colour plotted are differential colours 
calculated within rings corresponding 
to the standard apertures used for WISE photometry, i.e. 0-5.5, 5.5-8.25, 
8.25-11.0, 11.0-13.75,
13.75-16.5, 16.5-19.25, 19.25-22 and 22-24.75 arsec. 
As can be seen, all of the 19 galaxies have reasonably
high S/N differential W1-W2 colour measurements
over 4 or more standard aperture points i.e. out to at least 13.75
arcsec. We note that the optical radii of these galaxies are at least 2 times the
size of the PSF for both the W1 and the W2 bands, so
these colour profiles should be minimally affected by the small
difference between the W1 and W2 PSF (Wright et al 2010). 
A  W1-W2 colour of 0.8, marking the proposed boundary for
``pure'' AGN proposed in Stern et al (2013) is shown as a dotted
line in each panel.  As can be seen, all 19 galaxies have central (5
arcsecond aperture) colours that lie above this cut. As we will discuss in
the next section, the central torii of Seyfert galaxies are now known to
have physical sizes of a few parcsec.  If most of the hot dust emission
originates from a central torus, then the W1-W2 colours might be expected
to drop as a function of radius. Instead, we see that the W1-W2 colours
quite often {\em increase} with radius over the first two
aperture points,  and sometimes  remain above 0.8 out to
the edge of the galaxy. This implies that if red W1-W2 colours are caused by
emission from hot dust,  this dust must be spread over large radii (in some
cases even out to 10 kpc and beyond). The only galaxy where the W1-W2
colours drop significantly from the central 5.5 arcsec annulus to the
second 8.25 arcsec annulus is the very blue irregular galaxy, where the
central aperture is clearly centered on the compact star cluster. We thus
conclude that even the central aperture mid-IR emission from these galaxies
cannot be unambiguously associated with reprocessed radiation from an
accretion disk.

\subsection{ What is the best predictor for centrally peaked emission in
the W2 band?}

The analysis in the previous subsection of the resolved W1-W2 colour
profiles of the largest, most nearby galaxies with very red central colours
indicates that centrally peaked W2 emission is quite rare. In this section,
we ask whether we are able to locate a galaxy property that can be
identified as the {\em best predictor} for W1-W2 colours that are centrally
peaked.

To do this, we relax our selection criterion to include all galaxies with
$z<0.15$ where the error on the colour difference between the central 5.5
arcsecond and 8.25 arcsecond bins is less than 0.08. his
boundary was chosen iteratively to be a compromise bewtween sample size
and the accuracy of the colour gradient measurement. We also include all
galaxies with central 5.5 arcsec colours that lie above the 95th percentile
cut (i.e. this analysis is no longer restricted to the very reddest
outliers). This results in a sample of 996 galaxies. In Figure 10 we plot
the colour difference as a function of the following properties: stellar
mass, central stellar velocity dispersion, stellar surface mass density,
D$_n$(4000), Balmer decrement and radio luminosity.  The computed
correlation coefficient is indicated at the top of each panel. As can be
seen, the colour difference does not correlate with stellar mass or with
stellar surface density. It correlates weakly with velocity dispersion,
Balmer decrement and D$_n$(4000). The strongest correlation (by a factor of
two) is with radio luminosity.

\begin{figure}
\includegraphics[width=91mm]{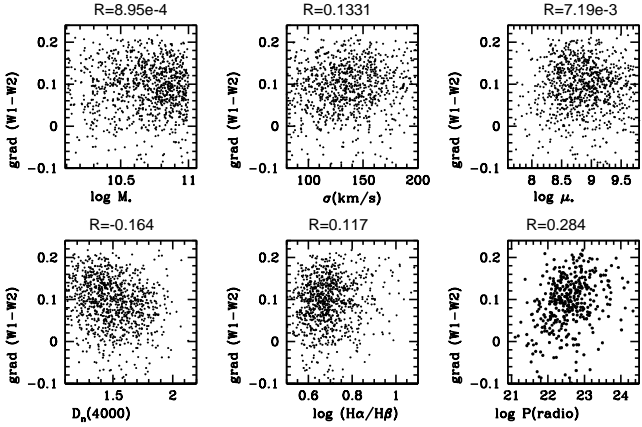}
\caption{ The colour difference between the central 5.5
arcsecond and 8.25 arcsecond bins  is plotted as a function of
stellar mass, central stellar velocity dispersion, stellar surface mass density,
D$_n$(4000), Balmer decrement and radio luminosity. Correlation
coefficients are indicated at the top of each panel.
\label{models}}
\end{figure}

In Figure 11, we colour-code the points in the colour difference versus
radio luminosity plane by D$_n$(4000) (left) and by central velocity
dispersion (right). For clarity, only galaxies in the
lowest and highest quartiles of the full sample are plotted. 
Clear separations are seen between the red and blue
points in both panels, which indicates that additional parameter dependences
in addition to radio luminosity are important in the prediction of the
colour gradient. Radio-loud galaxies with old stellar populations have
weaker colour gradients than radio loud galaxies with younger stellar
populations. This is easily understood if galaxies with low D$_n$(4000) have
ongoing star formation that is spread throughout a disk. More intriguing is
the finding shown in the right-hand panel that radio loud galaxies with
high central velocity dispersions have stronger W1-W2 colour gradients. We
note that  central stellar velocity dispersion is often used as a proxy for
black hole mass  (e.g. Heckman \& Kauffmann 2004), 
because there is a relatively tight correlation between
the two quantities in galaxies with  black hole masses that are accurately
measured using very high resolution stellar kinematical data. We
now ask whether the result shown in the right-hand panel of Figure 11 may
provide some tentative indication that black hole growth is favoured in the
bulges of radio-loud galaxies with centrally peaked mid-IR emission.

\begin{figure}
\includegraphics[width=91mm]{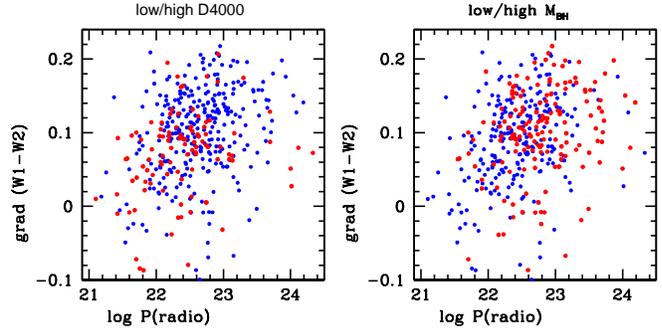}
\caption{The points in the colour difference versus
radio luminosity plane are colour-coded by D$_n$(4000) (blue,
low D$_n$(4000); red, high D$_n$(4000)) in the left-hand 
panel and by central velocity
dispersion (blue,low $\sigma$; red, high $\sigma$). 
For clarity, only galaxies in the 
lowest and highest quartiles of the full sample are plotted.
\label{models}}
\end{figure}

To help answer this question, we investigate the radio and optical
morphologies of these systems.  Figures 12 and 13 show optical $gri$ colour
and VLA FIRST radio images of a dozen radio-loud galaxies included in the
samples plotted in Figures 10 and 11.  We have selected galaxies with radio
luminosities greater than $10^{23.5}$ Watts Hz$^{-1}$ and we show images of
the 12 objects located at the lowest redshifts. The optical images show
that nearly all the galaxies are either clear merging systems or disturbed,
with asymmetric light profiles, double nuclei, outer shells and tidal arms.
The radio emission in all the VLA FIRST cutout-images is unresolved,
indicating that the radio emission originates from a compact region at the
center of the galaxy.  One galaxy (the merging system with clear double
nuclei and tidal features in the top right panel) has two radio nuclei. We
note that the VLA FIRST images have a resolution of 5 arcseconds and the
nearest galaxy in our sample is at a redshift of z=0.037. We are thus
unable to identify jets that are smaller than few kpc, so we cannot use the
radio images alone to distinguish between a central starburst occurring on
bulge scales, and a very compact relativistic jet.

\begin{figure}
\includegraphics[width=90mm]{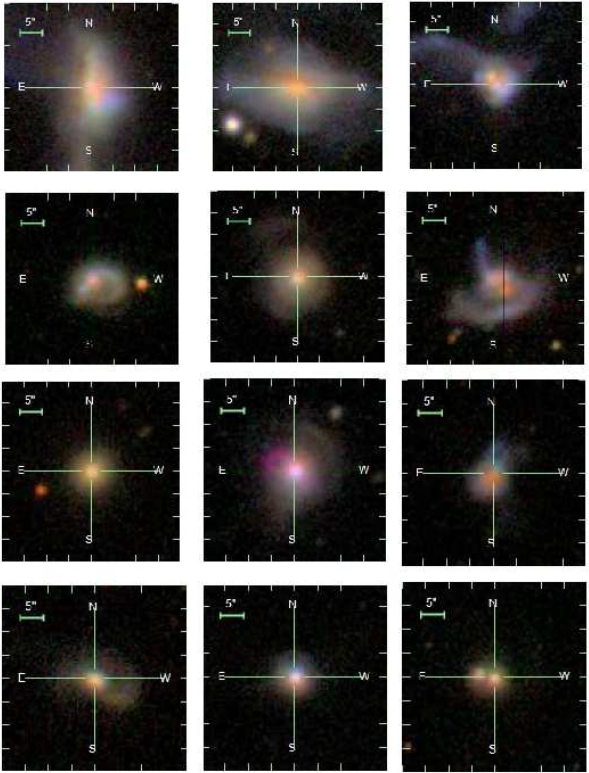}
\caption{ SDSS g,r,i colour cut-out images for a dozen
of the nearest red outliers with radio
luminosities greater than $10^{23.5}$ Watts Hz$^{-1}$.
\label{models}}
\end{figure}

\begin{figure}
\includegraphics[width=90mm]{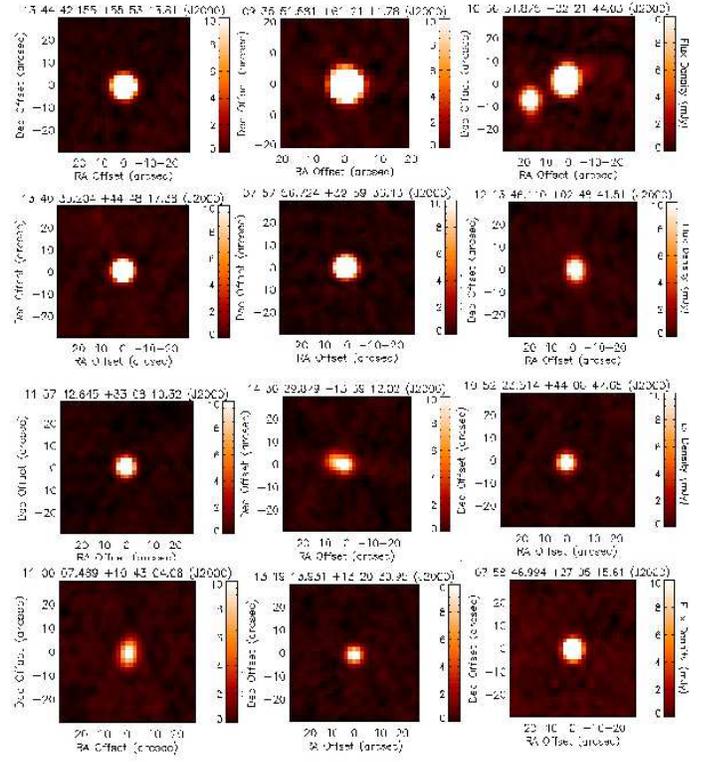}
\caption{ VLA cut-out images of the same galaxies shown in Figure 12, with
same orientation and image scale. 
\label{models}}
\end{figure}

As shown in Figure 1 of Kauffmann, Heckman \& Best (2008), $10^{23.5}$
Watts Hz$^{-1}$ marks the boundary in radio luminosity between galaxies
classified as star-forming using BPT (Baldwin, Phillips, \& Terlevich 1981) 
 emission line diagnostics, and
galaxies with LINER or Seyfert-like  emission lines.  Kauffmann, 
Heckman \& Best (2008) proposed a cut in the plane of radio-luminosity versus
extinction corrected H$\alpha$ luminosity in order to separate radio-loud
AGN from starburst galaxies. In Figure 14, we plot the location of all red
outliers with radio luminosities greater than $10^{23.5}$ Watts Hz$^{-1}$
and $z<0.15$ in the plane of radio luminosity versus extinction corrected
H$\alpha$ luminosity.  For reference, we plot all galaxies from the SDSS
main sample with $z<0.1$, $\log SFR/M_* > -11$ and which are not classified
as AGN in the BPT diagram as black points. As can be seen the radio-loud
red outliers all lie well above the main locus occupied by star-forming
galaxies. This provides additional support to our hypothesis that black
hole growth is occurring in these systems.

\begin{figure}
\includegraphics[width=75mm]{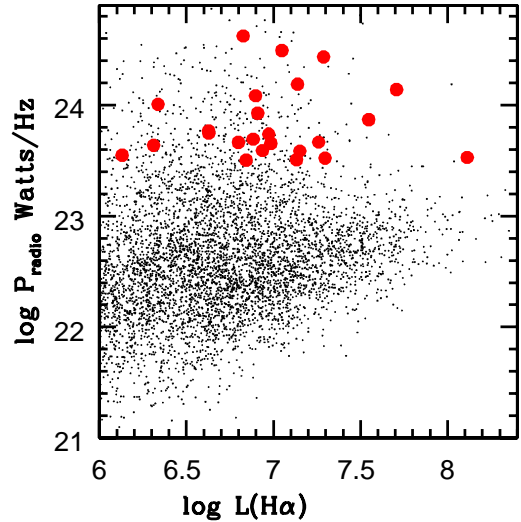}
\caption{The location of all red
outliers with radio luminosities greater than $10^{23.5}$ Watts Hz$^{-1}$
and $z<0.15$ in the plane of radio luminosity versus extinction corrected
H$\alpha$ luminosity.  For reference, we plot all galaxies from the SDSS
main sample with $z<0.1$, $\log SFR/M_* > -11$ and which are not classified
as AGN in the BPT diagram as black points.  
\label{models}}
\end{figure}

\section {Summary and Discussion}

The main observational findings presented in this paper are the following:
\begin {enumerate}

\item At fixed stellar mass $M_*$, the galaxy population as a whole
occupies a tight locus in the plane of W1-W2 colour versus W2 magnitude. We
show that the mean age of the stellar population as measured by the 4000
\AA\ break (D$_n$(4000)) decreases monotonically for redder W1-W2 colours
and higher W2 luminosities.

\item Optical AGN identified via BPT emission line ratio diagnostics occupy
the region of parameter space with bluer W1-W2 colours and smaller
luminosities in the W2 band.

\item At fixed stellar mass $M_*$ and $D_n(4000)$ , the distribution of
W1-W2 colours of the galaxy population is peaked around a small range of
values, but there is a long tail to much redder W1-W2 colour. We call this
tail the ``red outlier'' population.

\item We introduce a simple procedure to pull out the red outlier
population based on a combination of three stellar population diagnostics:
1) the 4000 \AA\ break strength, 2) the specific star formation rate
SFR/$M_*$, c) the H$\delta_A$ Balmer absorption line index, and apply this
to galaxies in the stellar mass range $10.1 < \log M_* < 11.2$. The red
outliers are more likely to be found in galaxies with higher stellar masses,
and are offset towards lower stellar surface densities,  
younger stellar
ages and higher dust content (as measured by the Balmer decrement
H$\alpha$/H$\beta$) compared to the optical AGN.  Twice as many of the red outliers are detected in the
radio as optical AGN. 

\item We examine W1-W2 colour profiles for a subset of the nearest, reddest
outliers with central W1-W2 colours greater than 0.8. This colour has been
proposed as a ``robust'' cut capable of identifying AGN at all redshifts by Stern et
al (2013).  Out of 19 galaxies, only one shows a clearly centrally peaked
colour profile. The W1-W2 colours in the other systems only dip below 0.8
at radial distance of greater than 5 kpc on average, indicating that the
hot dust emission is spread throughout the galaxy.

\item For a larger subset of $\sim$1000 of the nearest red outliers, we are able to 
look at correlations between the colour difference between W1-W2
measured within the central ($< 5$ arcsec) and 2nd (5-8 arcsec) WISE
photometric apertures, and a range of different galaxy properties. We find
that radio luminosity is the quantity that is most predictive of a redder
central W1-W2 colour.

\item We examine a subset of 20 of the nearest red outliers with radio
luminosities in excess of $10^{23.5}$ Watts Hz$^{-1}$. Their SDSS images reveal
that nearly all are morphologically disturbed galaxies with tidal features,
asymmetric light distributions and outer shell-like structures, or are systems
still in the process of merging. Their VLA images reveal compact,
unresolved emission at 1.5 Ghz in all but one of the merging systems, where
two compact radio sources are found.

\item We show that all red W1-W2 outliers with radio luminosities in excess 
of $10^{23.5}$ Watts Hz$^{-1}$ are also outliers in the relation between radio luminosity 
and extinction corrected H$\alpha$ luminosity, indicating that some fraction of
the radio emission is likely to originate 
from synchrotron emission from a relativistic jet. 80\%  percent of
such systems are identifiable as AGN based on optical emission line ratios
(see next section).

\end {enumerate}

\subsection{Relation to past work}

\subsubsection{Normal galaxies}

The result of our analysis that we found most surprising  was the large population of
normal  galaxies with very red W1-W2 colours with extended hot dust. 

We note that evidence for extended hot dust emission in  elliptical 
galaxies has existed for many years. 
Xilouris et al (2004) presented mid-infrared (MIR) maps for a sample of 18
nearby early-type galaxies observed at 4.5, 6.7 and 15 $\mu$m with the
ISOCAM instrument on board the ISO satellite. They modeled the broad-band
spectral energy distributions (SED) of these galaxies using  stellar
evolutionary synthesis models and  derived the  MIR excess over the stellar
component.  The morphology of the galaxies at these wavelengths was
examined  by plotting the azimuthally averaged radial profiles as well as
MIR colour profiles.  These authors found very diverse MIR proprties for
their sample.  Out of 18 galaxies studied, 2 dwarf ellipticals exhibited  a
clumpy distribution of dust, 2 galaxies exhibited  large scale disks, 2
galaxies were devoid of dust, 1 had compact MIR emission due to an AGN, 1
had emission associated with a jet, 3 had MIR emission which was more
centrally concentrated and the remaining 7 galaxies had smooth
distributions of the dust throughout the galaxy.

Kaneda et al 2007 analyzed observations of 7 galaxies, which were  among
the IRAS dusty elliptical galaxies described in Goudfrooij and de Jong
(1995). The dust masses derived for these objects from the IRAS flux
densities exceed by more than one order of magnitude the threshold where
dust is replenished by stellar mass loss.  For each galaxy, Kaneda et al
obtained 5--14 $\mu$m spectra using the IRS/Short-Low module in the standard
staring mode.  The main result was that in the outskirts of the galaxies,
the extended mid-IR emission appeared to follow the plasma distribution
rather than the stellar distribution for the three X-ray-brightest galaxies
in the sample. They proposed that  the spatial correspondence between mid-IR
and X-ray not only indicated the existence of diffusely distributed
interstellar dust, but also suggested the dominance of hot electron
collisions over stellar radiation for stochastic heating of distributed
dust (Draine \& Salpeter 1979; Dwek 1986).

The analysis in this paper indicates that extended hot dust is not confined
to the elliptical galaxy population. The red outliers
are found  over a large range in stellar mass and morphological type.

\subsubsection {Active galaxies and starbursts}

The mid-IR properties of  starburst galaxies, Seyfert galaxies and quasars  
have traditionally been studied  separately, but there
have been interesting  attempts to link
between different classes of objects.

The  existence of a IR enhancement or bump in the wavelength region
2-10 micron above the underlying power-law spectrum has been known in quasars since
the 1980s (Neugebauer 1976; Robson et al 1986; Edelson \& Malkan 1986). The
favoured model is that the near-IR excess originates from dust heated by
ultraviolet radiation from the central AGN to a temperature of 1500 K at a
radius of about 1 pc from the central source (e.g. Barvainis 1987).
Emission at wavelengths longer than 2 micron comes from cooler grains
farther from the central source.

Recently, infrared interferometry has made it possible  to resolve the
nuclear dust distributions in the dusty torii in nearby active galactic
nuclei.  In the Seyfert 2 galaxy NGC 1068, the interferometric
observations reveal a hot, parsec-sized disk that is surrounded by warm
dust extended in the polar direction (Wittkowski et al. 2004; Jaffe et al.
2004; Weigelt et al. 2004). Tristram et al (2013) report on observations of
the Circinius galaxy with the mid-infrared interferometric instrument (MIDI)
at the Very Large Telescope Interferometer. The dust emission is
distributed in two distinct components: a parsec-scale disk-like emission
component and a dusty outflow with the same temperature as the disk.
More recently, submillimetre observations 
of the torus of NGC 1068 using ALMA have 
provided a best-fit gas mass of $3 \times 10^5 M_{\odot}$ and radius of 3.5 pc
(García-Burillo et al 2016).

Nearby Seyfert galaxies also often host compact radio sources that can be
associated with either central star formation or with a jet. Gallimore et
al (2006) conducted a high-resolution VLA survey of 43 Seyfert and LINER
galaxies and found that 44\% (19 out of 43) show extended radio structures
at least 1 kpc in total extent. The radio emission does not match the
morphology of the disk or its associated star-forming regions. These
systems  also stand out by deviating significantly from the
far-infrared/radio correlation for star-forming galaxies.  Gallimore et al
propose that Seyfert galaxies generate radio outflows over a significant
fraction of their lifetime and propose a scenario in which virtually all of
the jet power is lost to the ISM within the galaxy.

Finally, the study of the most luminous galaxies in the far-IR, which are
predominantly merging systems, revealed a new class of star formation in
the central regions of these systems.  The most IR-bright starbursts with
an IR luminosity of $10^{11} M_{\odot}$ or more  have  characteristic sizes
of only 100 pc, with about $10^9 M_{\odot}$ of molecular gas contained
within the emitting region.  These compact starbursts have also been
studied at radio wavelengths. Out of 40 ULIRGS in the IRAS catalogue,
Condon et al (1991) find 25 objects with diffuse radio emission obeying the
far IR-radio correlation for normal starbursts. The other 15 have very
compact radio-loud nuclei -- in one object the radio emission is variable
and is constrained to be coming from region less than 1pc in diameter.

In summary, these more detailed studies support the view that galaxies with
centrally peaked mid-IR emission are those where black hole growth may be
occurring in a mode that is largely hidden at optical wavelengths, and
that black hole growth may be modulated/regulated by energetic feedback from
relativistic jets generated by the accreting black hole.

\subsection { Implications of extended hot dust emission in galaxies} 

The main goal of modern theories of galaxy formation is to incorporate a
broad range of observational phenomenology within a predictive model that
is based on evolving the distribution of dark matter and gas in the early
Universe forward in time using semi-analytic models or direct N-body
simulations.  We now view galaxies and their supermassive black holes  as
continuously evolving systems, which grow through gas accretion and merging
as their surrounding dark matter halos assemble. According to theory,  gas
accretion and merging in an individual galaxy is a stochastic process,
sometimes involving slow and steady accretion of new material onto the
disk, and sometimes involving major merging events that destroy disks, form
bulges and channel large amounts of gas into a central starburst (e.g.
Croton et al 2006).

In order for these theoretical models to reproduce key observational
constraints, such as the fraction of gas that is eventually turned into
stars in the central galaxies of the dark matter halos of $10^{12} M_{\odot}$
or greater, strong energy input from AGN is often invoked. Observational
evidence supporting this ``AGN feedback'' picture has thus far been
confined to direct measurements of outflowing gas in samples of a few dozen
luminous AGN (e.g. Greene et al 2011; Veilleux et al 2013). 
In this paper, we have presented evidence that the majority
of the 4.6 $\mu$m  mid-IR ``red outlier population", with W1-W2 colours
that are too red to be explained by their measured star formation rates,
are systems with spatially extended hot dust distributions. At distance of
5-10 kpc from the center of a galaxy, it is likely that the dust is being
heated by collisions with electrons from a hot plasma that permeates the
galaxy.  The question then arises as to why extended dust emission is so
common.  In regions of the galaxy where the volume filling factor of hot
plasma is high, interstellar dust is expected to be easily destroyed
through sputtering by ambient plasma ions (Draine \& Salpeter 1979; Dwek \&
Arendt 1992; Tielens et al. 1994).  There should thus be some physical
constraints on the {\em local} dust mass, which is determined by the
balance between destruction and replenishment.

One source of replenishment is mass loss from evolved AGB stars located in
the same region of the galaxy. Another possibility is the dust forms in the
central region of the galaxy during a starburst, and that outflows (perhaps
driven by jets) spread the dust out to large galacto-centric distances.
Spatially resolved dust formation and destruction mechanisms have recently
been introduced in cosmological hydrodynamical simulations (McKinnon,
Torrey \& Vogelsberger 2016).  In these simulations, supernovae driven
winds lead to dust being spread over scales of 100 kpc  around galaxies.
The predicted dust surface densities are too low to match observations of
dust in the circumgalactic medium around galaxies and the dropoff in
surface density is also too steep in comparison with observational
constraints from M\'enard et al (2010), suggesting that additional dust spreading
mechanisms may be needed.

\begin{figure}
\includegraphics[width=75mm]{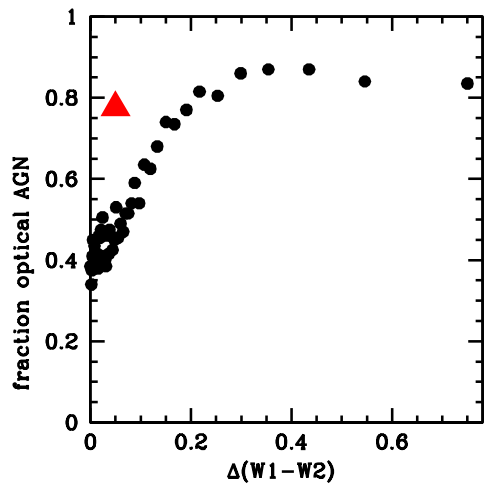}
\caption{ The black filled circles show the fraction of optical AGN in the red outlier
population as a function of $\Delta$(W1-W2), the difference between the W1-W2
colour of the galaxy and the colour that
delineates the 95th percentile cut. The optical AGN fraction drops from
15-20\% at the 95th percentile boundary, to a few percent for the very
reddest systems. The bright red triangle shows the optical AGN fraction for
all outliers with $\Delta$(W1-W2)$>$0.05 and with radio luminosities
greater than $10^{23}$ Watts Hz$^{-1}$.
\label{models}}
\end{figure}

\subsection {Future perspectives: what is the relation between the compact
radio-emitting sources and black hole growth?} We note that the
results presented in this paper show that if surveys of galaxies at mid-IR
wavelengths are to be used to study how black holes are formed and fed,
galaxies with compact hot-dust emission need to be separated from systems
where the mid-IR emission is emitted over large spatial scales.  The WISE
imaging data, with its low spatial resolution, is clearly not ideal for
identifying compact starbursts, but the combination of a  mid-IR colour cut
with detection of a compact radio-emitting source is likely to yield
samples that may be interesting for future follow-up at higher resolution.

The final issue we address is whether  the red  outliers are disjoint from the Seyfert 2
and LINERS selected using BPT diagram optical emission line ratio
diagnostics. Figure 15 answers this question. 
The black filled circles show the fraction of optical AGN in the red outlier
population as a function of $\Delta$(W1-W2), the difference between the W1-W2 
colour of the galaxy and the colour that
delineates the 95th percentile cut. The optical AGN fraction increases  from
40\% at the 95th percentile boundary, to 85\%  for the very
reddest systems. The bright red triangle shows the optical AGN fraction for
all outliers with $\Delta$(W1-W2)$>$0.05 and with radio luminosities
greater than $10^{23}$ Watts Hz$^{-1}$. We find that 77\% of these
radio-loud red outliers are classified as optical AGN, which is similar to the fraction
found for the very reddest outliers in W1-W2 colour. 
This result is not sensitive to the adopted
threshold in radio luminosity. We conclude that mid-IR  selection does not reveal
a significant population of AGN that cannot be identified optically.  
In future work we will study the optical emission and stellar absorption line properties
of the radio-loud, mid-IR outlier population in more detail. 


\end{document}